\documentclass[12pt]{article}
\usepackage{graphicx}
\usepackage{cite}
\usepackage{color}
\begin{document}
\title{The spectrum of scalar-meson nonets\\
in the Resonance-Spectrum Expansion\footnote{Talk presented at the
workshop on "Scalar Mesons and Related Topics"
honoring the 70th birthday of Michael Scadron}
}
\author{
Eef van Beveren\\
{\normalsize\it Centro de F\'{\i}sica Te\'{o}rica,
Departamento de F\'{\i}sica, Universidade de Coimbra}\\
{\normalsize\it P-3004-516 Coimbra, Portugal}\\
{\small http://cft.fis.uc.pt/eef}\\ [.3cm]
\and
George Rupp\\
{\normalsize\it Centro de F\'{\i}sica das Interac\c{c}\~{o}es Fundamentais,
Instituto Superior T\'{e}cnico}\\
{\normalsize\it Universidade T\'{e}cnica de Lisboa, Edif\'{\i}cio Ci\^{e}ncia,
P-1049-001 Lisboa, Portugal}\\
{\small george@ist.utl.pt}\\ [.3cm]
{\small PACS number(s): 14.40.Ev, 14.40.Lb, 14.40.Cs, 13.25.-k}
}
\maketitle
\begin{abstract}
We argue that the low-lying scalar-meson nonet
\cite{PRD26p239} makes part
of a subset of a family of infinitely many scalar-meson nonets,
which in turn makes part of a family
of infinitely many quark-antiquark bound states and resonances.
We outline the properties of this subset.
\end{abstract}
\maketitle
\section{Introduction}
Except for a few mesons, like pions and kaons,
most quark-antiquark states show up as resonances
in systems of two or more mesons.
It is thus opportune to study the interplay of meson-meson scattering
and $q\bar{q}$ confinement
\cite{PRD21p772,ARXIV07112080}.

The mesonic resonances extracted from experiment are organized by
flavor content, $J^{PC}I^{G}$ quantum numbers, mass, and width.
From the few hundred listed in the PDG~tables \cite{JPG33p1},
one would not yet conclude that they are abundant.
Nevertheless, based on the $b\bar{b}$ and $c\bar{c}$ spectra,
we concluded in Ref.~\cite{PRD21p772} that there must exist
an infinity of such states,
though cut off from observation at higher masses
because of the many two-meson systems coupling to $q\bar{q}$.
Accordingly, we expect an infinity of scattering poles to show up
in meson-meson scattering, here represented by
\begin{equation}
E\; =\; P_{0}\, ,\;\; P_{1}\, ,\;\; P_{2}\, ,\;\;\dots
\;\;\; .
\label{polepositions}
\end{equation}
Unitarity then requires that in the one-channel restriction,
assuming the poles (\ref{polepositions})
to be simple poles,
the elastic scattering matrix $S$ be given by\footnote{
Note that we do not consider here
a possible overall phase factor
representing a background.}
\begin{equation}
S(E)
\; =\;
\frac{\textstyle\raisebox{5pt}{
$\left( E-P_{0}^{\ast}\right)\left( E-P_{1}^{\ast}\right)
\left( E-P_{2}^{\ast}\right)\dots$}}
{\textstyle\raisebox{-5pt}{
$\left( E-P_{0}\right)\left( E-P_{1}\right)\left( E-P_{2}\right)\dots$}}
\;\;\; .
\label{unitairS}
\end{equation}

If we assume that the resonances (\ref{polepositions})
stem from an underlying confinement spectrum,
given by the real quantities
\begin{equation}
E\; =\; E_{0}\, ,\;\; E_{1}\, ,\;\; E_{2}\, ,\;\;\dots
\;\;\; ,
\label{confinementpositions}
\end{equation}
then we may represent the differences $\left( P_{n}-E_{n}\right)$,
for $n=0$, 1, 2, $\dots$, by $\Delta E_{n}$.
Thus, we obtain for the unitary $S$-matrix the expression
\begin{equation}
S(E)
\; =\;
\frac{\textstyle\raisebox{5pt}
{$
\left( E-E_{0}-{\Delta E_{0}}^{\ast}\right)
\left( E-E_{1}-{\Delta E_{1}}^{\ast}\right)
\left( E-E_{2}-{\Delta E_{2}}^{\ast}\right)
\dots$}}
{\textstyle\raisebox{-5pt}
{$
\left( E-E_{0}-\Delta E_{0}\right)
\left( E-E_{1}-\Delta E_{1}\right)
\left( E-E_{2}-\Delta E_{2}\right)
\dots$}}
\;\;\; .
\label{unitairSdeltaE}
\end{equation}

So we assume here that resonances occur in scattering because
the two-meson system couples to confined states, usually
of the $q\bar{q}$ type, viz.\ in non-exotic meson-meson scattering.
Let the strength of the coupling be given by $\lambda$.
For vanishing $\lambda$, we presume that the widths and real shifts
of the resonances also vanish. Consequently, the scattering poles
end up at the positions of the confinement
spectrum~(\ref{confinementpositions}), and so
\begin{equation}
\lim_{\lambda\downarrow0}\,\Delta E_{n}\;=\;0
\;\;\;\;{\textstyle \mbox{\rm for}}\;\;\;\;
n\; =\; 0,\;\; 1,\;\; 2,\;\;\dots
\;\;\; .
\label{smalllambda}
\end{equation}
As a result, the scattering matrix tends to unity,
as expected in case there is no interaction.

An obvious candidate for an expression
of the form (\ref{unitairSdeltaE})
looks like
\begin{equation}
S(E)
\; =\;
\left[
1+\lambda^{2}\,\left\{{\displaystyle \sum_{n}^{}}
\frac{\textstyle G(E)^{\ast}}{\textstyle E-E_{n}}\right\}
\right]\;\left[
1+\lambda^{2}\,\left\{{\displaystyle \sum_{n}^{}}
\frac{\textstyle G(E)}{\textstyle E-E_{n}}\right\}
\right]^{-1}
\;\;\; ,
\label{unitairSRSE}
\end{equation}
where $G$ is a smooth complex function of energy $E$.
\clearpage

\section{Kaon-pion S-wave scattering}
In order to compare expression (\ref{unitairSRSE})
with results of experiment,
we must choose a suitable complex function $G$.
This has been done in Ref.~\cite{PRD21p772},
and was further developed in
Refs.~\cite{PRD27p1527,PRL91p012003,PRL93p202001,PRL97p202001}.
Furthermore, values for the real spectrum (\ref{confinementpositions})
must be chosen.
In principle, one could fit $E_{n}$ ($n=0$, 1, 2, ...) to experiment.
But it is our experience that the spectrum listed in Ref.~\cite{JPG33p1}
is not yet rich enough to determine a suitable confinement spectrum.
In Refs.~\cite{PRD21p772,PRD27p1527,PRL91p012003,PRL93p202001,PRL97p202001}
we proposed flavor masses and a universal level spacing $\omega$
for this purpose.
The latter quantity can, with some confidence, be deduced from the
$c\bar{c}$ and $b\bar{b}$ spectra,
and also from the light positive-parity resonances \cite{QNP06p91}.
One finds $\omega =0.19$ GeV, corresponding to
interquark distances ranging from 0.2 fm for $b\bar{b}$
to 0.6 fm for light quarks..

\begin{figure}[htbp]
\begin{center}
\begin{tabular}{ccc}
\includegraphics[height=120pt]{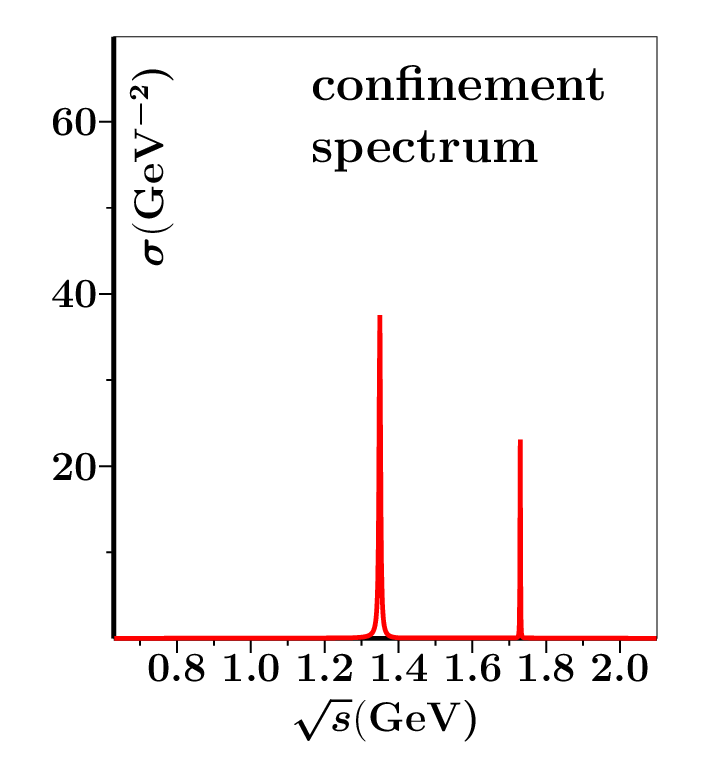} &
\includegraphics[height=120pt]{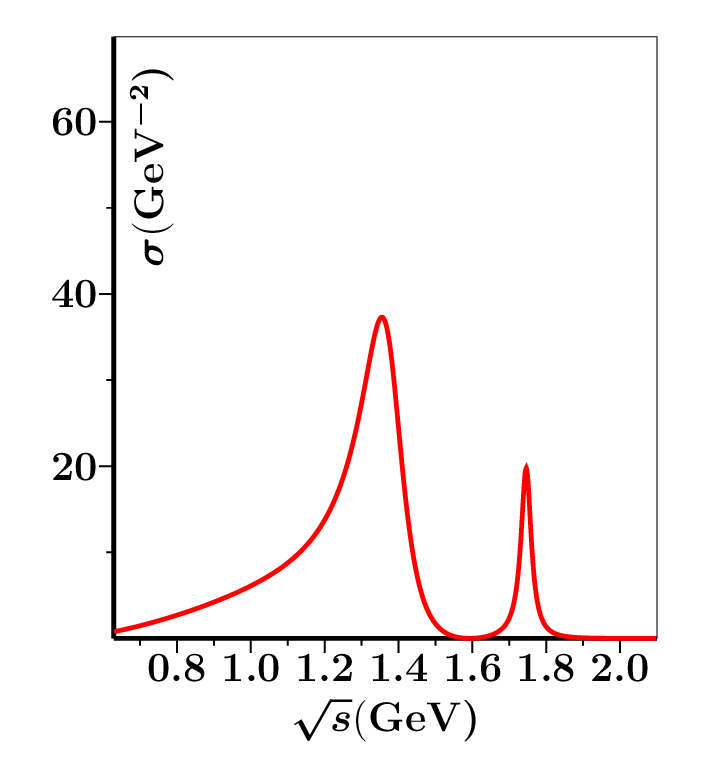} &
\includegraphics[height=120pt]{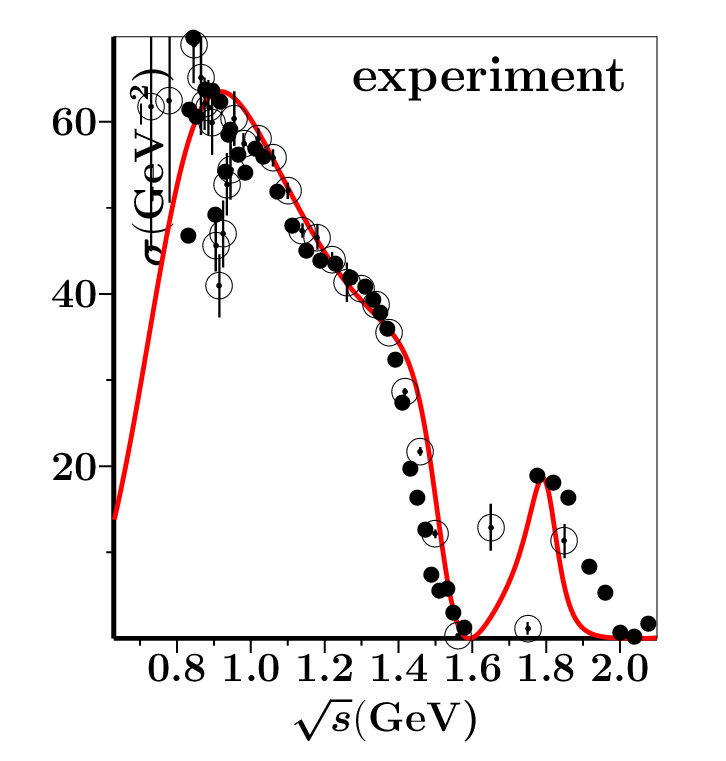}\\ [-5pt]
(a) & (b) & (c)
\end{tabular}
\end{center}
\caption{Cross section for $S$-wave isodoublet $K\pi$ scattering.
Left: for very small values of $\lambda$, one observes
the $J^{P}=0^{+}$ $n\bar{s}$ confinement spectrum.
Middle: when $\lambda$ takes about half its model value, one notices
some more structure for low invariant masses.
Right: at the model value of $\lambda$, the latter structure becomes
dominant and well in agreement with the experimental observations.
The data are taken from Ref.~\cite{NPB133p490} (open circles)
and Ref.~\cite{NPB296p493} (full circles).
}
\label{KpiS}
\end{figure}

After fitting the parameters to heavy-heavy, heavy-light and light-light
vector and pseudoscalar data \cite{PRD27p1527},
we turn our attention to the scalar mesons
\cite{ARXIV07111435}.
This is just a matter of setting quantum numbers $J^{P}=0^{+}$,
determining $E_{0}$, calculating cross sections from
expression (\ref{unitairSRSE}),
and comparing to available data.
Here, we will concentrate on the nonstrange-strange ($n\bar{s}$)
centers of gravity of the scalar nonets, for energies up to about 2 GeV,
and use the elastic-scattering data of Refs.~\cite{NPB133p490,NPB296p493}.

In Fig.~\ref{KpiS} we show how cross sections
following from formula~(\ref{unitairSRSE}) vary with increasing $\lambda$,
for $S$-wave isodoublet $K\pi$ scattering.
In Fig.~\ref{KpiS}a the $n\bar{s}$ confinement
spectrum is well visible for small $\lambda$,
whereas in Fig.~\ref{KpiS}c, for the model value of $\lambda$,
experiment is reproduced.
We find a fair agreement for total invariant masses up to 1.6 GeV.
\clearpage

\begin{figure}[htbp]
\begin{center}
\begin{tabular}{ccc}
\includegraphics[height=130pt]{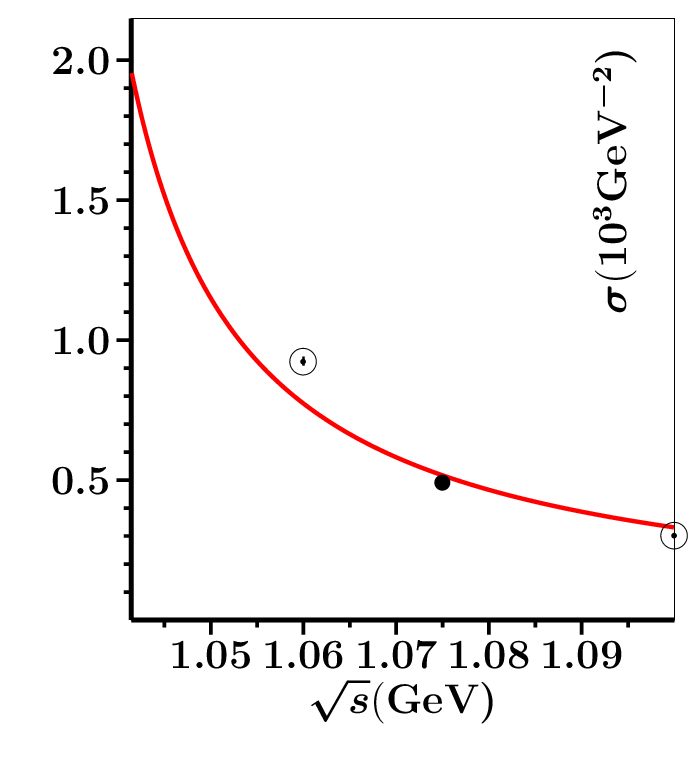} &
\includegraphics[height=130pt]{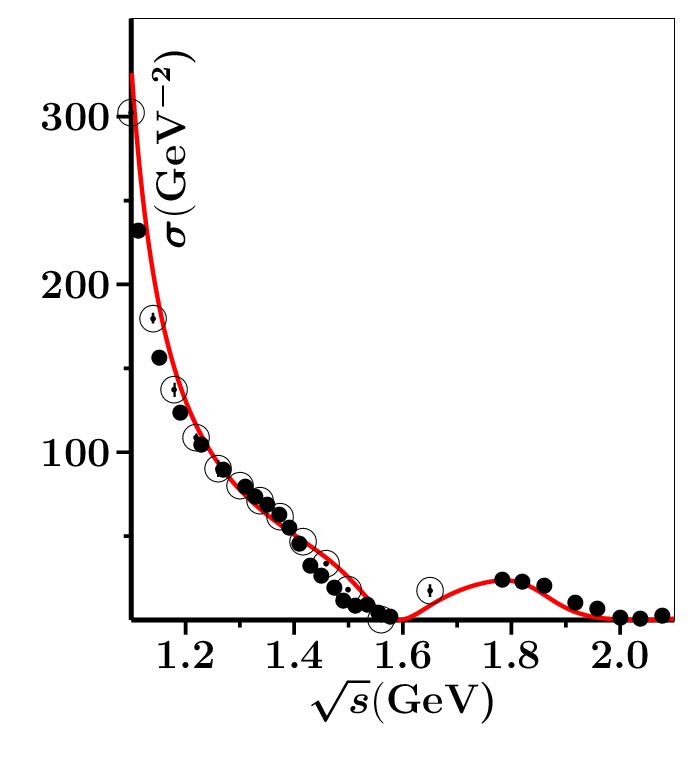} &
\includegraphics[height=130pt]{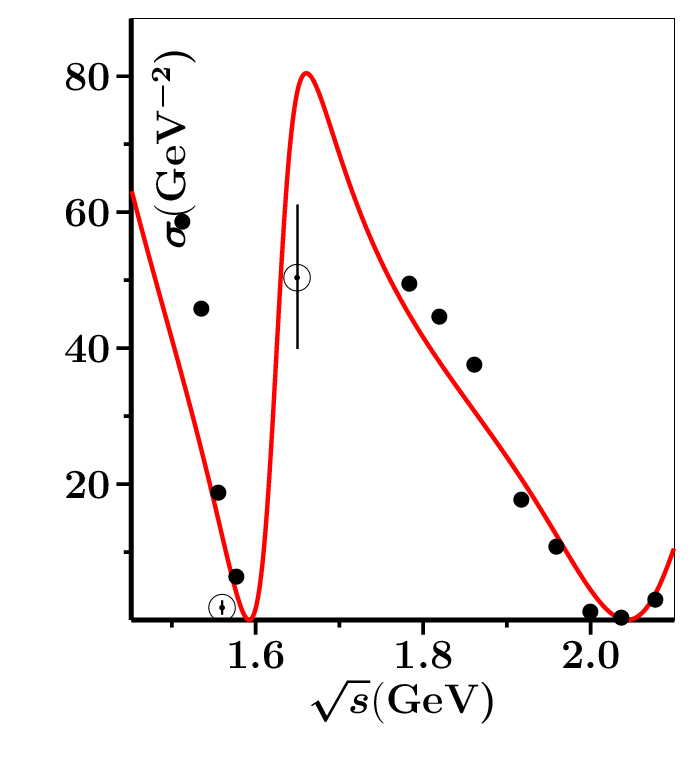}\\ [-5pt]
(a) & (b) & (c)\\ [-125pt]
$K\eta$ & \hspace{40pt}$K\eta$ & \hspace{60pt}$K\eta'$\\ [115pt]
\end{tabular}
\end{center}
\caption{$S$-wave $K\eta$ and $K\eta'$ ``cross section'' (see text),
as a function of total invariant mass.
({\bf a}): $K\eta$ from threshold up to 1.1 GeV.
({\bf b}): $K\eta$ from 1.1 GeV up to 2.1 GeV.
({\bf c}): $K\eta'$ from threshold up to 2.1 GeV.
The data are taken from Ref.~\cite{NPB133p490} (open circles)
and Ref.~\cite{NPB296p493} (full circles).
}
\label{KetaS}
\end{figure}
Now, in order to have some idea about the performance of formula
(\ref{unitairSRSE}) for $S$-wave $I\!=\!1/2$ $K\pi$ scattering
at higher energies, we argue that,
as in our model there is only
one non-trivial eigen-phase shift for the
coupled $K\pi$+$K\eta$+$K\eta'$ system, we may compare the phase shifts
of our model for $K\eta$ and $K\eta'$ to the experimental phase shifts
for $K\pi$.  We do this comparison in Fig.~\ref{KetaS},
where, instead of the phase shifts, we plot the cross sections,
assuming no inelasticity in all cases.
We observe an extremely good agreement.
In particular, for $K\eta'$ (Fig.~\ref{KetaS}c)
we become aware of a structure in the data at about 1.9 GeV,
indicating the presence of a not anticipated pole.
This is something we would not have easily noticed from the data alone.
\clearpage

\section{A basketful of scalar nonets}
When we inspect formula (\ref{unitairSRSE}) for poles
in the $S$-wave isodoublet $K\pi$ scattering amplitude,
then we find the pole structure as summarized in Table~\ref{KpiSpoles},
i.e., five poles at energies up to about 2.2 GeV real part.
The first pole, at $0.772 - 0.281 i$ GeV,
describes the $K^{\ast}_{0}(800)$ structure \cite{ARXIV07121605},
whereas the second pole, at $1.52 - 0.097 i$ GeV, represents the
well-established $K^{\ast}_{0}(1430)$ resonance \cite{EPJC52p55}.
\begin{table}[htbp]
\begin{center}
\begin{tabular}{||c|c|c|c|c|c||}
\hline\hline
Pole (GeV) & $0.772 - 0.281 i$ & $1.52 - 0.097 i$ &
$1.79 - 0.052 i$ & $2.04 - 0.15 i$ & $2.14 - 0.065 i$\\
Origin & continuum & confinement & confinement & continuum & confinement\\
\hline\hline
\end{tabular}
\end{center}
\caption{$T$-matrix poles for $S$-wave $K\pi$ scattering,
as obtained from Eq.~(\ref{unitairSRSE}).}
\label{KpiSpoles}
\end{table}

Our model is explicitly flavor independent, meaning that the only flavor
breaking in formula~(\ref{unitairSRSE}) stems from the effective quark masses,
which determine the ground state of the confinement spectrum
(\ref{confinementpositions}), and from the masses of the mesons in the
scattering channels.
Consequently, $\pi\pi$ scattering is not very
different from $K\pi$ scattering in our model.
We may expect then that each of the two flavor combinations
that couple to isoscalar $S$-wave
$\pi\pi$ and $KK$ scattering has a pole structure similar to the one
in isodoublet $K\pi$ scattering, with the proviso that
$n\bar{n}$-$s\bar{s}$ mixing in the $I\!=\!0$ case introduces an extra
complication \cite{ARXIV08021209}.
Hence, since also $\eta\pi$ is similar to $K\pi$ in our model
\cite{HEPPH0207022},
with each pole of Table~\ref{KpiSpoles} we associate
a full nonet of scalar mesons.
The often read comment that {\it too many isoscalar states are observed}
\/\cite{PLB541p22}, in order to justify the application of alternative quark,
or even quarkless, configurations \cite{IJMPA20p5156}, is not confirmed here.
\clearpage

\section{\bf Confinement and continuum poles}
In order to explain the difference between
{\it confinement} \/and {\it continuum} \/poles (see Table~\ref{KpiSpoles}),
we turn to another member of the scalar-meson family,
namely the heavy-light ($c\bar{s}$) $D_{s0}^{\ast}(2317)$ meson
\cite{PRC72p065202}.

The mass of the $D_{s0}^{\ast}(2317)$ ends up below the threshold
of the lowest OZI-allowed decay mode (i.e., $DK$).
Consequently, it represents a bound state
in this specific selection of decay channels \cite{HEPPH0610327},
which we consider the most important.
Accordingly, the $D_{s0}^{\ast}(2317)$ may be represented
by a pole on the real energy axis.
The pole representing the first radial excitation of the $c\bar{s}$
system in a relative $P$-wave comes out well above the $DK$ threshold.
In Ref.~\cite{PRL97p202001}, two poles were found,
one at 2.32 GeV and a second at $(2.85-i0.024)$ GeV,
representing the ground state and the first radial
excitation of the $J^{P}=0^{+}$ $c\bar{s}$ system, respectively.
Experiment \cite{HEPEX0607082} reported a $c\bar{s}$ structure at 2.86 GeV,
with the same line-shape as our theoretical prediction
\cite{PRL97p202001},
and being compatible with $J^{P}=0^{+}$ quantum numbers.

But in Ref.~\cite{PRL97p202001} an additional pole showed up
in the scattering amplitude.
Its theoretical position was reported at $(2.78-i0.23)$ GeV.
\begin{figure}[htbp]
\begin{center}
\begin{tabular}{l}
\resizebox{0.25\textwidth}{!}{\includegraphics{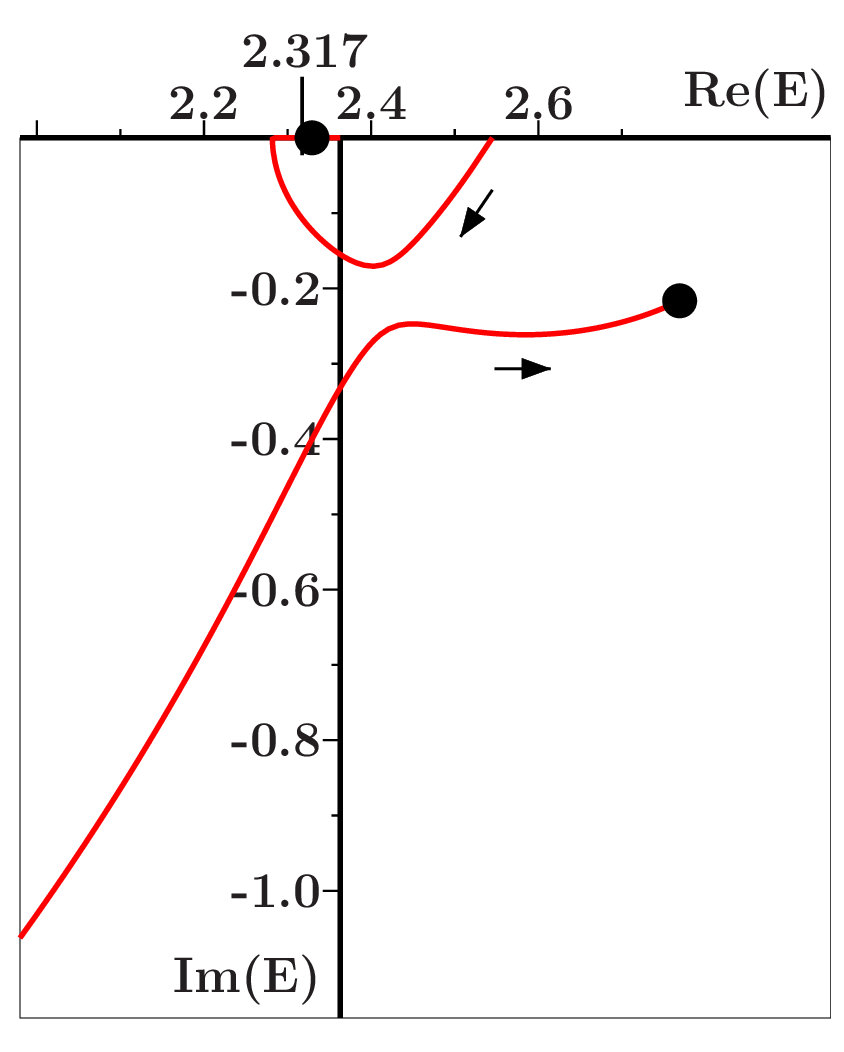}}
\end{tabular}
\end{center}
\caption{Trajectories of poles in the $DK$ $S$-wave
scattering amplitude for the model
of Ref.~\cite{PRL97p202001}, as a function
of the amount of unquenching.
In the quenched approximation, the dynamically generated pole has negative
infinite imaginary part, whereas the confinement ground state comes out at
$\sqrt{s}=2.454$ GeV on the real axis.
The arrows indicate how the poles move when unquenching increases.
The model's physical values are indicated by dots.
The imaginary axis is drawn at the $DK$ threshold.}
\label{Ds0poles}
\end{figure}
In Fig.~\ref{Ds0poles} we show the trajectories of
the two lowest-lying poles in the scattering matrix
for increasing $c\bar{s}$-$DK$ coupling
\cite{PRL91p012003}.
The BABAR collaboration reported in Ref.~\cite{HEPEX0607082}
on the possible existence of a broad $c\bar{s}$ resonance,
which might correspond to the dynamically generated pole
\cite{PRD77p014012}.

Expression (\ref{unitairSRSE}) thus yields more poles
than we bargained for\footnote{Recently,
related studies heve been carried out in
Refs.~\cite{PRD70p114013}
}.
The jump from Eq.~(\ref{unitairSdeltaE}) to Eq.~(\ref{unitairSRSE})
contains the physics outlined in Ref.~\cite{PRD21p772}:
meson pairs with non-exotic quantum numbers couple to $q\bar{q}$ states
through $^{3}P_{0}$ quark-pair annihilation/creation.
This mechanism yields the resonances which we expected from the
quark-antiquark confinement. But it also yields
quasi-bound meson-meson molecules
due to shielding caused by the
quark-pair annihilation/creation.
By model-reducing the intensity of the latter process,
the associated poles move into the continuum and disappear
from the spectrum of resonances in meson-meson scattering
(see Fig.~\ref{KpiS}).
The low-lying scalar mesons belong to this set of resonances
\cite{PLB652p250}.

The two trajectories shown in Fig.~\ref{Ds0poles} come close
to each other for certain values of the $c\bar{s}$-$DK$ coupling.
Upon a variation of one other model parameter, this becomes a saddle point.
Depending on the value of this parameter, the trajectories may interchange.
In that case the end points are connected differently,
making the $D_{s0}^{\ast}(2317)$ the dynamically generated state,
whereas the other pole then seems to stem from the confinement ground state.
This is actually what appears to happen for the light positive-parity
ground-state mesons and makes them move up in energy when unquenching is
turned on.
For the scalar mesons, these states correspond to the
$f_{0}(1370)$, $f_{0}(1500)$, $K_{0}^{\ast}(1430)$, and $a_{0}(1450)$.
The dynamically generated poles correspond
to the nonet of lower-lying scalar mesons \cite{ZPC30p615}.
\clearpage

\section{Conclusions}
Most probably, mesons are just mixtures
\cite{IJMPA20p5156,CS86p379}
of
{\it quark-antiquark} \/states,
{\it two-meson molecules},
{\it glueballs},
{\it tetraquarks},
{\it hexaquarks},
{\it hybrids},
and so forth.
We have shown
that the first two of the latter list of possible
components are the most relevant ones
\cite{PRD21p772,PRD27p1527,PRL91p012003,PRL93p202001,PRL97p202001}.
Moreover, a resonance is really a collection of states,
all with different masses.
Each of these states will have a different composition.

In the spectrum of scattering poles for
two-meson systems coupled to $q\bar{q}$ states,
we find an infinity of resonances
consisting of two distinguishable subsets.
One subset manifests the phenomenon of $q\bar{q}$ confinement,
whereas the other subset is a direct consequence of quark-pair
annihilation/creation.
The ground-state scalar nonet of confinement poles
is formed by the nonet
$f_{0}(1370)$, $f_{0}(1500)$, $K_{0}^{\ast}(1430)$, and $a_{0}(1450)$.
On the other hand, the low-lying scalar nonet
$f_{0}(600)$, $f_{0}(980)$, $K^{\ast}_{0}(800)$, and $a_{0}(980)$
is the lowest-in-mass scalar nonet of continuum poles.
\clearpage

\section*{Acknowledgements}
We thank the
{\it Funda\c{c}\~{a}o para a Ci\^{e}ncia e a Tecnologia}
of the {\it Minist\'{e}rio da
Ci\^{e}ncia e do Ensino Superior} \/of Portugal
for financial support,
under contract POCI/FP/\-63907/2007.

\newcommand{\pubprt}[4]{#1 {\bf #2}, #3 (#4)}
\newcommand{\ertbid}[4]{[Erratum-ibid.~#1 {\bf #2}, #3 (#4)]}
\def\AIPCP{AIP Conf.\ Proc.}
\def\CS{Curr.\ Sci.}
\def\EPJC{Eur.\ Phys.\ J.\ C}
\def\IJMPA{Int.\ J.\ Mod.\ Phys.\ A}
\def\JPG{J.\ Phys.\ G}
\def\MPLA{Mod.\ Phys.\ Lett.\ A}
\def\NPB{Nucl.\ Phys.\ B}
\def\PLB{Phys.\ Lett.\ B}
\def\PRC{Phys.\ Rev.\ C}
\def\PRD{Phys.\ Rev.\ D}
\def\PRL{Phys.\ Rev.\ Lett.}
\def\PREP{Phys.\ Rept.}
\def\PTPS{Prog.\ Theor.\ Phys.\ Suppl.}
\def\ZPC{Z.\ Phys.\ C}

\end{document}